\documentclass[
  prl,aps,showpacs,twocolumn,preprintnumbers,nofootinbib,
  amsmath,amssymb,superscriptaddress,longbibliography,
  10pt,floatfix
]{revtex4-2}

\usepackage[T1]{fontenc}
\usepackage[english]{babel}

\usepackage{graphicx}
\usepackage{comment}

\usepackage{xcolor}        
\usepackage[normalem]{ulem}
\usepackage{soul}

\usepackage{amsmath,amssymb,amsfonts}
\usepackage{cancel}
\usepackage{chngcntr}
\usepackage{braket}
\usepackage{nicefrac}

\usepackage{upgreek}
\usepackage{chemmacros}
\chemsetup{greek=upgreek}
\usepackage{siunitx}
\usepackage{nameref}
\usepackage[colorlinks=true,linkcolor=red,citecolor=blue,urlcolor=blue]{hyperref}

\newcolumntype{C}{>{$}c<{$}}

\DeclareSIUnit{\dBm}{dBm}

\begin{document}
\title{Compact self-matched gyrators using edge magnetoplasmons}

\author{Aldo Tarascio}
\email{aldo.tarascio@unibas.ch}
\affiliation{Department of Physics, University of Basel, Klingelbergstrasse 82, CH-4056, Basel, Switzerland}

\author{Yiqi Zhao}
\affiliation{Department of Physics, University of Basel, Klingelbergstrasse 82, CH-4056, Basel, Switzerland}

\author{Rafael S. Eggli}
\affiliation{Department of Physics, University of Basel, Klingelbergstrasse 82, CH-4056, Basel, Switzerland}

\author{Taras Patlatiuk}
\affiliation{Department of Physics, University of Basel, Klingelbergstrasse 82, CH-4056, Basel, Switzerland}

\author{Christian Reichl}
\affiliation{Laboratory for Solid State Physics, ETH Zürich, 8093 Zürich, Switzerland} 
\affiliation{Quantum Center, ETH Zürich, 8093 Zürich, Switzerland} 

\author{Werner Wegscheider}
\affiliation{Laboratory for Solid State Physics, ETH Zürich, 8093 Zürich, Switzerland} 
\affiliation{Quantum Center, ETH Zürich, 8093 Zürich, Switzerland} 

\author{Stefano Bosco}
\affiliation{Department of Physics, University of Basel, Klingelbergstrasse 82, CH-4056, Basel, Switzerland}
\affiliation{QuTech and Kavli Institute of Nanoscience, Delft University of Technology, Delft, Netherlands}

\author{Dominik M. Zumbühl}
\email{dominik.zumbuhl@unibas.ch}
\affiliation{Department of Physics, University of Basel, Klingelbergstrasse 82, CH-4056, Basel, Switzerland}

\begin{abstract}

Edge magnetoplasmons provide a natural platform for chiral electrodynamics, where broken time-reversal symmetry enforces unidirectional propagation. When probed at microwave frequencies, they offer a route to compact non-reciprocal devices. So far, implementations have suffered from large losses or required complicated matching networks. Here we show that the circulating modes coupled to capacitive gates give rise to a gyrator response, characterized by directional $\pi$ phase difference between forward and reverse transmission. By engineering a three-terminal capacitive geometry, we realize a self-impedance matched gyrator in which the gyration points coincide with transmission maxima, enabling nearly lossless gyration without external matching networks. Our devices are implemented on a GaAs 2D gas, operate from 0.2 to \qty{2}{\giga\hertz}, tuned by magnetic field, with sub-millimeter footprints and insertion loss as low as \qty{2}{\deci\bel}. This is a factor of 100 smaller and less lossy than commercial and plasmon units, respectively.
A dissipative model, in agreement with experiment, provides the fundamental physics and delivers the key materials parameters, leading the way to even less lossy devices approaching ideal operation by materials improvement.
The self-impedance matched concept is broadly applicable to a variety of devices, thus providing a foundation for a new generation of high-quality microwave plasmon technology.

\end{abstract}

\maketitle

\section{Introduction}
Edge magnetoplasmons (EMPs) are collective charge oscillations propagating chirally along the boundary of a two-dimensional electron gas (2DEG) and have been extensively studied in GaAs~\cite{Allen1983,Volkov1988,Aleiner1994,Ashoori1992,Grodnensky1991,Wassermeier1990,Talyanskii1992,Talyanskii1994}.
Such EMPs can be investigated by looking at the microwave response of 2DEG systems and used for compact non-reciprocal devices~\cite{Viola2014,Bosco2017,Bosco2017a,Bosco2025,Mahoney2017}, given their intrinsic chiral propagation under a perpendicular magnetic field that allows passive, magnetic-field-tunable operation from a few GHz to sub-GHz. Moreover, the EMP planform has been explored for coherent long-distance coupling and control of semiconductor spin qubits~\cite{Elman2017,Bosco2019,Bosco2019a,Bartolomei2023,Lin2024,Lin2026}, highlighting its potential as a versatile resource for quantum information processing.

Modern cryogenic RF and quantum information systems rely on microwave components that enforce directionality in signal propagation in order to suppress noise back-action on quantum devices~\cite{Clerk2010,Bardin2021}. Conventional ferrite-based components, such as non-reciprocal phase shifters, circulators, and isolators, become impractically large when designed to operate in the few-hundred-MHz range~\cite{Linkhart2014,Dunn1965} and are therefore challenging to integrate on-chip in large-scale quantum processors~\cite{Sliwa2015,Linkhart2014}.
Beyond conventional microwave signal routing, non-reciprocity has been shown to play a critical role in controlling energy and information flow in quantum thermodynamics~\cite{Ahmadi2024,Zou2024}, with a growing recognition of non-reciprocity as a versatile resource for quantum technologies~\cite{Barzanjeh2025}.

A fundamental non-reciprocal device is the gyrator~\cite{Tellegen1948,Hogan1953}, which transmits microwave signals in one direction while imparting a $\pi$ phase shift in the opposite direction. Gyrators are key elements in microwave engineering, with applications in impedance conversion, signal routing, and multiplexing~\cite{Pozar2012,Linkhart2014}. Gyrators have been proposed as key building blocks for quantum error correction, for instance by enabling bosonic encodings such as Gottesman–Kitaev–Preskill codes to realize self-correcting qubits~\cite{Cai2021,Rymarz2021,Brady2024}. Moreover, as a non-reciprocal phase shifter by integrating a gyrator in a Mach-Zehnder interferometer~\cite{Hiyama2015,Bosco2017,Singh2022,Pan2024} it can be used to build circulators and isolators, commonly used in many cryogenic setups~\cite{Chen2012,Abdo2019,Krinner2019,Bardin2021,HarveyCollard2022,Oppliger2025}.

Previous EMP-based non-reciprocal devices typically required external matching circuits and suffered from large insertion losses~\cite{Mahoney2017,Martinez2025}. We are able to overcome the insertion loss limitations by realizing and characterizing a self-impedance matching gyrator scheme proposed in Ref.~\cite{Bosco2017}. 
We employ a three-terminal device in which the potentials of two electrodes are measured relative to a third grounded electrode that is twice as long as the other two. This geometry ensures that a relative $\pi$ phase difference results between the two transmission directions at the resonant frequencies of the EMP modes. As a result, the device achieves self-impedance matching while showing gyration. We examine performance and frequency operation ranges and explain the experimental results with a dissipative theoretical model~\cite{Bosco2025} that includes the self-impedance matching port, extracting quantitative parameters such as dissipation.
Thanks to the self-impedance matching design we record low losses with a minimum of \qty{2}{\deci\bel} in a wide range of frequencies from GHz down to sub-GHz frequencies with footprint on the order of a millimeter, allowing for on-chip integration with other components in a scalable architecture.

\section{The Device}

\begin{figure}[htpb]
\centering
\includegraphics[width=0.85\linewidth]{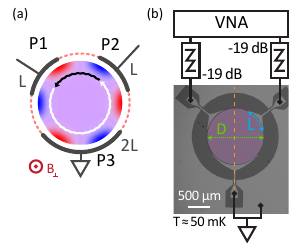}
\includegraphics[width=0.85\linewidth]{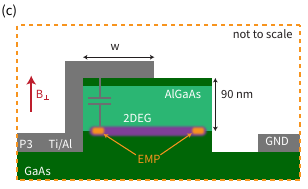}

\caption{\textbf{Device and measurement scheme.}
(a) Circuit model of the device.  A magnetic field  \( B_\perp \) is applied out-of-plane causing chiral propagation (white and black arrows) of the EMP resonant modes (red and white) along the edge. Port P3 is connected to ground. The pink dashed curves represent ungated sections.
(b) Optical micrograph and microwave measurement setup of the large device. The circular mesa hosting a high-mobility 2DEG is false-colored in purple, with diameter $D$ (green arrow) and the contact overlap length $L$ (blue arrow). Measurements are done at $\sim$\qty{50}{\milli\kelvin}. 
(c) Cross-sectional schematic along the dashed orange line in (b), showing the heterostructure (green and dark green) with the 2DEG (purple) located 90\,nm below the surface and an aluminum gate overlapping the mesa.}
\label{fig:Fig1}
\end{figure}

The gyrator device consists of a disc of 2DEG defined by etching, where a perpendicular magnetic field creates an EMP channel at the edge supporting resonant eigenmodes coupled to three ports P1-P3 arranged around the perimeter, shown in Fig.~\ref{fig:Fig1}. We present data from devices of different diameters $D$, \qty{1225}{\micro\meter} and \qty{780}{\micro\meter} referred to as the \emph{Large} and \emph{Small} devices. The ports are capacitively coupled with a few micrometer sized overlap $w$ on top of the 2DEG, as shown, removing the need for ohmic contacts and thus eliminating a possible source of dissipation. The port and ungated gaps between them are all of the same length $L$, except for one port, P3, which is twice as long and is kept grounded, following Bosco \emph{et al}. Ref. \cite{Bosco2017}. This provides a passive way for self-impedance matching and allows for low-loss operation.
An external magnetic field \( B_\perp \) defines the chirality of an edge magnetoplasmon propagating along the edges of the disc. This breaks reciprocity causing the forward transmission (white arrow) to take a different path than the reverse transmission (black arrow). While the present implementation relies on an applied magnetic field, similar devices can also be realized in anomalous Hall materials, where symmetry breaking enables magnetic-field-free operation~\cite{Mahoney2017a,Martinez2025}.
Despite being grounded, the third port does not act as a loss channel for the microwave signal, since signal extraction would require a matched \qty{50}{\ohm} load.

Besides capacitive coupling, the metallic gates play another role: They slow down EMP propagation by screening long-range Coulomb interactions. This screening reduces the gated propagation velocity $v_{\rm g}$ by more than one order of magnitude, and can be estimated as~\cite{Zhitenev1994,Zhitenev1995,Kamata2010,Kumada2011}
\begin{equation}
   v_{\rm g} \approx \frac{\sigma_{0}}{c_{\rm emp}},
\end{equation}
where $c_{\rm emp}$ is the gate capacitance per unit length to the EMP and $\sigma_{0}$ is the conductivity amplitude~\cite{Bosco2025}.
The gate screening effect shifts device operation from several GHz down to the sub-GHz regime for devices on the scale of $\sim$\qty{1}{mm} at \qtyrange{50}{400}{\milli\tesla}.
For the high mobility heterostructure we are using and for the frequencies and fields investigated this conductance reduces to the Hall conductivity and we have $\sigma_{0}\approx\sigma_{xy}$. 
When operating the device, we see a reduction in velocity of more than an order of magnitude between gated and ungated sections.

\section{Microwave Response}

With the geometry defined, we characterize the device by measuring the magnitude and phase of the transmission parameters without low-temperature amplification.

\subsection{Phase}

\begin{figure}[htpb]
\centering
\includegraphics[width=\linewidth]{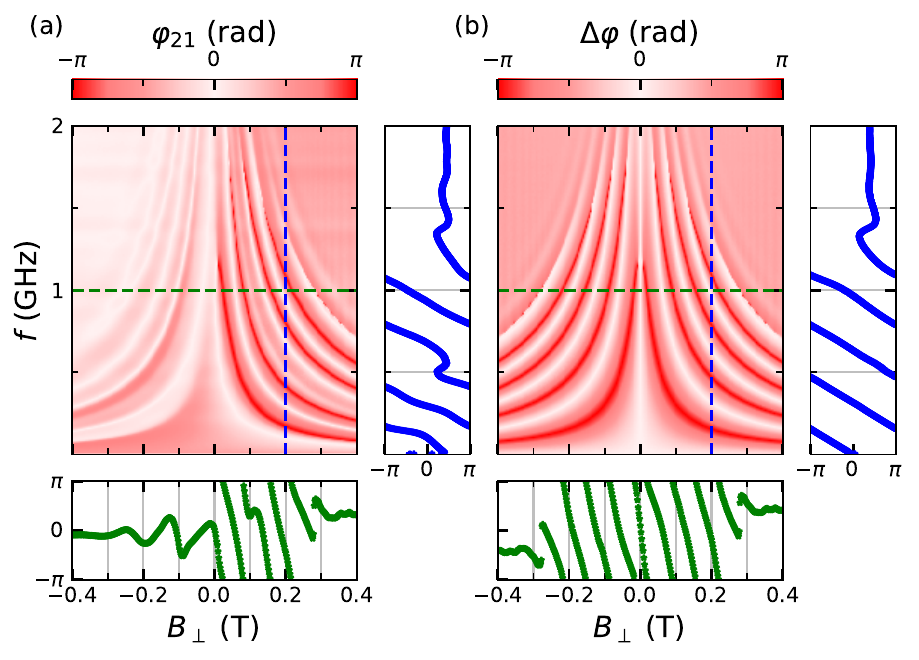}

\caption{\textbf{Phase response of the large device.}
(a) Phase of the forward transmission parameter \( S_{21} \) after subtraction of the electrical delay.
(b) Phase difference \( \Delta\varphi\) between forward and reverse transmission. The red curves indicate points of gyration. Horizontal and vertical cuts at fixed frequency and magnetic field are shown in the bottom and side panels as indicated by the dashed lines.}
\label{fig:Fig2}
\end{figure}

The hallmark of gyrator operation is its non-reciprocal phase response, imparting a $\pi$ phase shift in only one direction, therefore we carefully evaluate the phase response. The measurement setup introduces an electrical delay associated with the cable length~\cite{Johnson2011}. The phase response of the device can be retrieved once such a linear background is removed (see methods), defining the calibrated phase $\varphi_{21}$ shown in Fig.~\ref{fig:Fig2}(a), for the transmission from port 1 to port 2.

The phase response $\varphi_{21}$ is not symmetric in magnetic field, and the signal is imparted with a larger phase shift for positive magnetic fields, corresponding to the longer propagation path according to the white arrow in the diagram of Fig.~\ref{fig:Fig1}(a). In the fixed field cuts of Fig.~\ref{fig:Fig2}(a) (blue dashed line and trace), there is an almost linear phase accumulation. When the phase is mapped to $[-\pi , \pi]$ this results in apparent phase jumps. This is also visible as the bright red curves of Fig.~\ref{fig:Fig2}(a), where the color scale maps $\pi$ and $-\pi$ to the same red color.

The reverse propagation phase $\varphi_{12}$ (see Supplementary Information Fig.~S1) displays the same features for opposite magnetic fields.
Instead of relying on the setup-dependent electrical delay, which is difficult to calibrate accurately, we measure transmission in both directions and define the \emph{non-reciprocal phase difference}:
\begin{equation}
\Delta\varphi \equiv \mathrm{Arg}\left[ e^{i(\varphi_{21} - \varphi_{12})} \right],
\end{equation}
which inherently cancels the line delay and prevents the introduction of arbitrary offsets.
Some features seen in the cuts of Fig.~\ref{fig:Fig2}(a) were the result of reciprocal features from the setup transmission lines such as resonances, and therefore disappear in $\Delta\varphi$.
With this definition, $\Delta\varphi$ is wrapped to the interval $[-\pi , \pi]$, ensuring a consistent representation of the non-reciprocal phase and providing a clear and unambiguous signature of gyration when $\Delta\varphi \simeq \pm \pi$, as seen in the 4 bright and symmetric red curves in Fig.~\ref{fig:Fig2}(b). In the line cuts of Fig.~\ref{fig:Fig2}(b) it can be seen that now the phase winding is almost perfectly linear in field and frequency.  The winding rate can be tuned by changing $B_\perp$, resulting in several tunable $\pi$-phase difference points.  Deviations from the linear winding, such as a flattening of $\Delta\varphi$ at large magnetic fields and frequencies, are due to dissipation effects~\cite{Bosco2025}.

\subsection{Self-Impedance Matching}

An ideal gyrator transmits signals perfectly, free of any losses from both ports with a non reciprocal phase $\Delta\varphi=\pm\pi$. In practice, the frequency of the gyration points is set by the phase winding introduced by the path-length difference, but these points do not necessarily coincide with peaks in the transmission magnitude. In the self-impedance matched design, however, the gyration points align with specific transmission peaks, resulting in a narrow and tunable gyration bandwidth~\cite{Bosco2017,Bosco2025}.

\begin{figure}[htpb]
\centering
\includegraphics[width=1\linewidth]{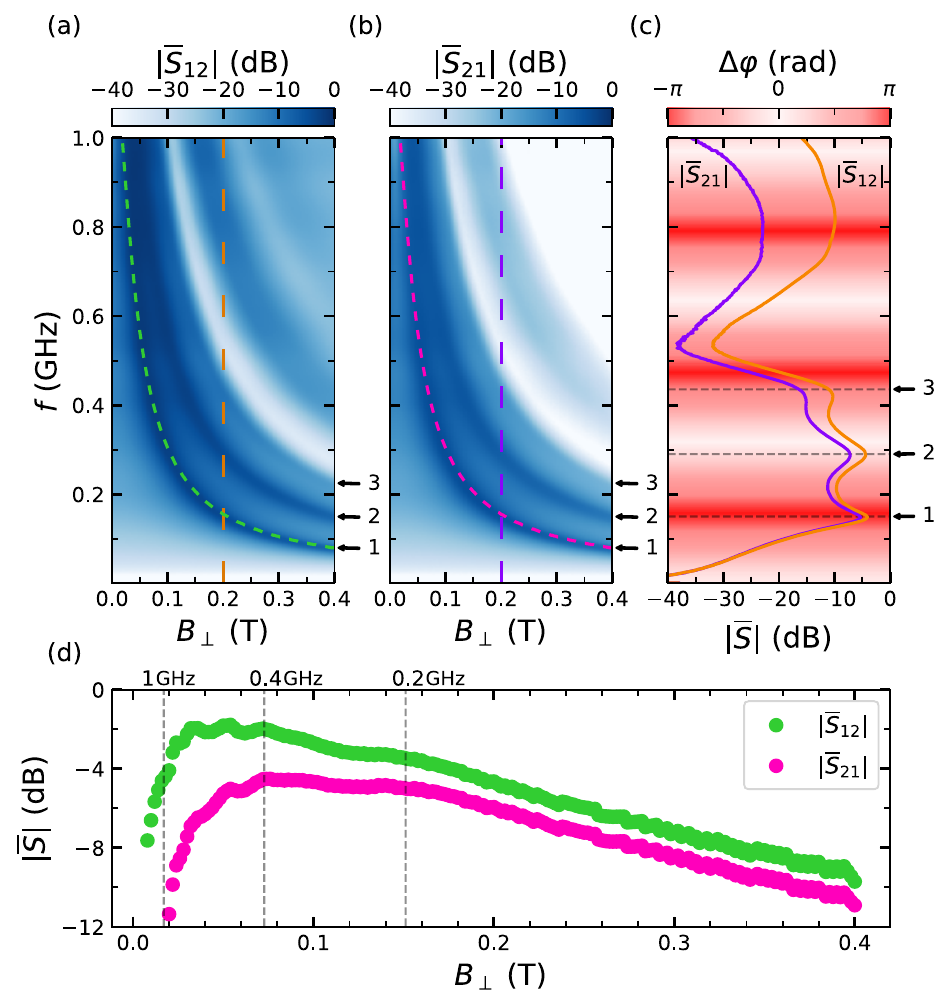}
\caption{\textbf{Magnitude response of the large device.} 
(a) and (b) Magnitude of reverse ($|\overline{S}_{12}|$) and forward ($|\overline{S}_{21}|$) transmitted signal. The green and magenta dashed curves represent the lowest-frequency gyration mode. The black arrows serve to number the three peaks, as labeled. 
(c) Field cuts at \qty{200}{\milli\tesla}, indicated by the orange and purple dashed lines in panels (a) and (b), sharing the same frequency axis. The background  shows $\Delta\varphi$ taken at the same field as the cuts. The black arrows and black dashed lines indicate the position of the numbered magnitude peaks, showing $\Delta\varphi \approx \pi$ for peaks 1 and 3, while $\Delta\varphi \approx 0$ for peak 2. 
(d) Insertion loss of peak 1, as indicated by the green and magenta dashed curves in panels (a) and (b). The gray dashed lines represent the corresponding points in frequency, showing a wide range of tunability. }
\label{fig:Fig3}
\end{figure}

To quantify the insertion loss, the measured transmission magnitudes were calibrated against a shorted sample holder measured under identical conditions, removing the effect of the setup. The calibrated reverse and forward transmission ($|\overline{S}_{12}|$ and $|\overline{S}_{21}|$) are shown in Fig.~\ref{fig:Fig3}(a,b) for positive magnetic fields. A characteristic three peak structure emerges above $\sim$\qty{50}{\milli\tesla}, labeled as peaks~1,~2, and~3. A clear asymmetry develops between the two transmission directions: the longer EMP propagation path ($|\overline{S}_{21}|$) exhibits stronger attenuation but, consistent with Fig.~\ref{fig:Fig2}(a), a larger phase delay. For negative fields, the situation reverses, as shown in the Supplementary Information.

As shown in Fig.~\ref{fig:Fig3}(c), the two outer peaks 1 and 3 correspond to a phase difference of approximately $\pi$, while the central peak 2 remains near zero phase difference. Hence, peaks 1 and 3 mark the gyration points of the device, where near-ideal non-reciprocal transmission occurs, while the central peak represents a symmetric transmission condition with no effective gyration. A slight misalignment between the magnitude peaks and the phase features arises from additional delay accumulated along the ungated path, and will be discussed in the model validation section. The same three peak behavior is captured by our model: when the dissipation is sufficiently small the simulation reproduces the observed three-peak structure (Supplementary Information for details). The lowest-frequency peak (peak~1) exhibits the smallest insertion loss, as shown in Fig.~\ref{fig:Fig3}(d), with insertion loss of \qtyrange[range-phrase=~or~]{2}{4}{\deci\bel} depending on the direction of propagation, reflecting the chiral and dissipative nature of EMP propagation in this regime.

\subsection{Size and non-reciprocity parameter}
Two device diameters were investigated: a large device ($D = \qty{1225}{\micro\meter}$, shown in the main figures) and a smaller one ($D = \qty{780}{\micro\meter}$, shown in the Supplementary Information). To compare their performance directly, we define a dimensionless non-reciprocity parameter
\begin{equation}
\Delta = \frac{|\overline{S}_{21} - \overline{S}_{12}|}{2},
\end{equation}
which can express the gyrator performance in one number being $\Delta = 1$ for an ideal gyrator with unit transmission and a $\pi$ phase difference between forward and reverse propagation. Any deviation in phase or magnitude reduces $\Delta$, making it a compact figure of merit of non-reciprocity and facilitates comparison across different devices and operating conditions.

\begin{figure}[htpb]
\centering
\includegraphics[width=\linewidth]{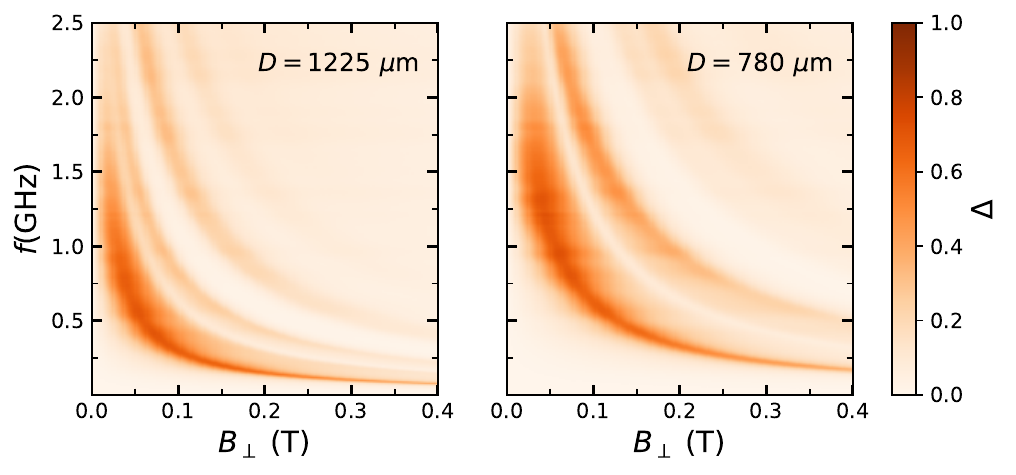}
\caption{\textbf{Non-reciprocity of two devices with different diameters.} 
Non-reciprocity parameter $\Delta$ as a function of $B_\perp$ and frequency for two gyrators with diameters $D$ as labeled. The gyration modes appear as dark peaks, reaching maximum performance of $\Delta = 0.70$ at \(B_\perp = \pm 70\,\text{mT}\) and \(f = 400\,\text{MHz}\) for the large device, and $\Delta = 0.72$ at \(B_\perp = \pm 62\,\text{mT}\) and \(f = 940\,\text{MHz}\) for the small device.}
\label{fig:Fig4}
\end{figure}

The measured $\Delta$ maps for the large and small devices are shown in Fig.~\ref{fig:Fig4}. Both devices achieve a maximum $\Delta$ of approximately 0.7, indicating strong non-reciprocity deviating from ideal behavior mostly due to dissipation, reducing the magnitude. The smaller device operates at higher frequencies, consistent with its reduced circumference and shorter EMP propagation path, and it also exhibits enhanced performance for the second gyrating peak compared to the larger device, likely due to the reduced loss along the shorter path.

\section{Model Validation}
The device response is captured by combining the self-impedance matching scheme introduced in Ref.~\cite{Bosco2017} with the dissipative stub model of Ref.~\cite{Bosco2025}. 
Physically, the model describes the capacitive coupling of the electrodes to chiral EMPs propagating along the boundary of the Hall conductor. The edge dynamics support a set of eigenmodes corresponding to circulating charge-density waves with discrete eigenfrequencies
\begin{equation}
\omega_m = m\omega_R,  
\label{eq:modes}
\end{equation}

where $\omega_R = \sigma_0/(cR)$ sets the fundamental EMP frequency scale determined by the conductivity $\sigma_0$, gate capacitance per unit length $c$, and device radius $R$. Each eigenmode corresponds to a traveling EMP circulating around the perimeter. The microwave response of the device arises from the excitation and interference of these edge modes by the electrodes.

Dissipation enters the model through the parameter \(\delta\), which measures the deviation from purely transverse Hall transport due to resistive losses. The limit \(\delta=0\) corresponds to dissipationless Hall transport, while \(\delta=\pi/2\) represents maximal dissipation and results in a reciprocal device. In the operating regime we find \(\delta \lesssim 0.1\).

\begin{figure}[htpb]
\centering
\includegraphics[width=0.85\linewidth]{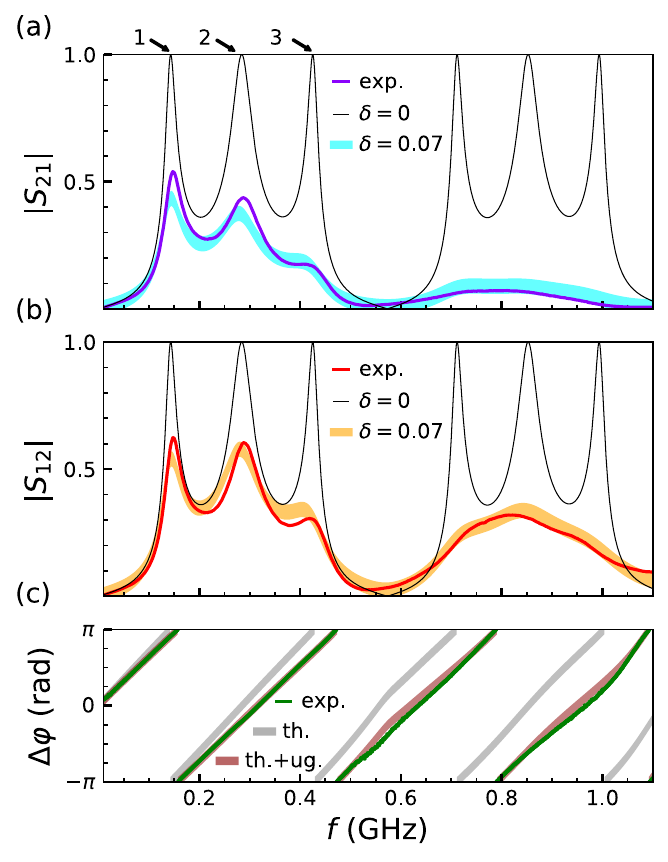}
\caption{\textbf{Model fitted to the large device.}
(a,b) Linear magnitude of $\overline{S}_{21}$ and $\overline{S}_{12}$ for the large device: experimental data at \qty{200}{\milli\tesla} (purple and red) compared with the fitted model (cyan and orange bands). The black line shows the dissipationless limit ($\delta = 0$). 
(c) Phase difference $\Delta\varphi$ of the large device: experimental data (green) compared with the model without additional delay (gray band) and with the linear shift introduced by an ungated delay $\tau_{\mathrm{ug}} \approx \qty{0.3}{\nano\second}$ (brown band).}
\label{fig:Fig5}
\end{figure}

We apply the model with two main simplifications. First, we account only for EMP propagation in the gated regions, where electrostatic screening slows the mode considerably~\cite{Volkov1988,Johnson2003,Kumada2011} and therefore dominates the response. Second, we assume purely edge conduction and neglect any bulk contribution. The model was fitted to data acquired at a fixed magnetic field in order to reproduce the experimental behavior. The agreement is remarkable: the simulated traces capture the three peak structure in the magnitude and the asymmetric transmission amplitudes, as shown in Fig.~\ref{fig:Fig5}(a,b). The comb of modes is due to the EMP eigenmodes in the disc described in Eq.~\ref{eq:modes}. The sub-structure of three peaks followed by a dip arises from the four-segment geometry of the terminals when neglecting the ungated sections and having a terminal twice the length of the other two. Peaks 1 to 3 correspond to phases of $\pi,2\pi,3 \pi$, respectively, given by the phase accumulated in the path length difference, mapping to $\Delta\varphi=\pi,0,\pi $. The dip originates from an overall net zero charge on the terminal, leading to total reflection of the signal (see Supplementary Information).
The phase difference in Fig.~\ref{fig:Fig5}(c) exhibits the expected linear winding with frequency, consistent with the measurements. In the absence of any additional delay, peaks 1 and 3 align with $\Delta\varphi=\pi$, while peak 2 aligns with $\Delta\varphi=0$. However, the phase difference predicted by the model $\Delta\varphi_{\rm th}$ neglects an additional phase $\Delta\varphi_{\rm ug}$ arising from propagation through the ungated sections of the device, given by $2\pi f \tau_{\mathrm{ug}}$, where $\tau_{\mathrm{ug}}$ is the propagation time in the ungated regions. Accounting for this contribution allows us to directly determine the EMP velocity in the ungated sections and on by using the resonant frequency coming from the magnitude peaks, which reflects the combined propagation through both gated and ungated regions, we can then also extract the gated EMP velocity.
We find a dependence on \(1/B\) for the EMP velocity both in the gated and ungated cases, as expected in the low field regime~\cite{Volkov1988,Johnson2003,Zhitenev1994}. We report the velocities in the Supplementary Information.

By fitting traces between \qtyrange[]{100}{400}{\milli\tesla}, we find that the dissipation parameter \(\delta\) initially follows the expected $1/B$ trend associated with Hall transport, but saturates above \( \sim\)\qty{200}{\milli\tesla}, approaching an approximately field-independent value of \(\delta \simeq 0.06\), as shown in the Supplementary Information. Even before the saturation sets in, the extracted dissipation exceeds expectations based on the wafer mobility and carrier density. This behavior suggests the presence of an additional loss mechanism that acts as an approximately constant longitudinal resistance independently of the field. The resulting finite dissipation attenuates the gyration peaks leading to increased losses at higher fields, where one would expect perfect self-impedance matching ~\cite{Bosco2017}.
Overall, the model reproduces the device behavior with a small set of parameters, enabling estimation of the EMP velocities and quantitative characterization of the losses.

\section{Conclusion}
We have demonstrated a compact cryogenic gyrator based on EMP propagation in a GaAs electron gas. By implementing the self-impedance matching scheme of Ref.~\cite{Bosco2017}, the device achieves non-reciprocity, insertion loss of few \qty{}{\deci\bel}, and sub-millimeter footprint dimensions, enabling integration in scalable on-chip architectures. By comparing devices of different sizes, we have shown how the operating frequency scales with geometry, in good agreement with the analytic model of Ref.~\cite{Bosco2025}. This model accurately reproduces the observed spectral features, such as the three peak structure and non-reciprocal phase, and provides a quantitative route to extract device parameters such as capacitances, velocities, and dissipation. 

Our results establish EMP-based non-reciprocal devices as a viable platform for microwave engineering in the sub-GHz to GHz range, overcoming the footprint and field compatibility limitations of ferrite based components.
The design principles in our device can then be used for optimizing signal strength and resonance conditions in key components for building scalable quantum interconnects and quantify their performance by looking at the dissipation.

Looking ahead, the concepts and methods presented here can be extended with gate tunability, as recently explored in related systems~\cite{Frigerio2024}, and adapted to other platforms with chiral edge states, such as silicon or germanium heterostructures or quantum anomalous Hall systems that do not require external magnetic fields fields to operate~\cite{Mahoney2017a,Gourmelon2023,Roeper2024,Martinez2024,Martinez2025}.
Beyond signal routing, EMP-based devices offer a platform for coupling, driving and entangling semiconductor qubits over long distances~\cite{Elman2017,Bosco2019,Bartolomei2023,Lin2024,Lin2026}, representing yet another interesting application.

\section{Methods}
\subsection{Material}
The GaAs/AlGaAs heterostructure hosts a 2DEG located $90\,\mathrm{nm}$ below the surface, with an electron density $n = 2.35\times 10^{11}\,\mathrm{cm^{-2}}$ and a mobility $\mu = 3.9\times 10^{6}\,\mathrm{cm^{2}/Vs}$, as measured by the grower at \qty{1.3}{\kelvin} on a separate chip.

Two devices of different mesa diameters were fabricated and measured and the parameters are reported in the table below.

\begin{table}[htpb]
\centering
\begin{tabular}{lccc}
\hline
 & diameter $D$ & length $L$ & overlap $w$ \\
\hline
Large  & \qty{1225}{\micro\meter} & \qty{550}{\micro\meter} & \qty{5}{\micro\meter} \\
Small  & \qty{780}{\micro\meter}  & \qty{350}{\micro\meter} & \qty{3.2}{\micro\meter} \\
\hline
\end{tabular}
\caption{Geometrical parameters of the fabricated devices.}
\end{table}

\subsection{Theoretical Framework}
We model the driving potential from Eqs.~(9)–(13) of Ref.~\cite{Bosco2025} and set the voltage response of the large grounded electrode to zero. By introducing a three‐terminal configuration and grounding one electrode we create a common reference for the other two ports. With this arrangement, the voltages at the two active ports are defined relative to the grounded electrode, rather than to each other. Combining the boundary conditions with the voltage response Eq.~(14a) of Ref.~\cite{Bosco2025} is derived, and we obtain the a Fourier series for the dimensionless admittance matrix:
\begin{equation}
Y_{\rm emp}(c_{\mathrm{emp}}, \sigma_0, \delta).
\end{equation}
The field dependence of the response is included implicitly through the parameters  \(c_{\rm emp}\),~\( \sigma_0\),and \(\delta\).
The frequency of the device is set by the characteristic frequency
\begin{equation}
\omega_R = \frac{2\sigma_{0}}{c_{\mathrm{emp}} D^*}.
\end{equation}
The first magnitude peak appears at $f=\omega_R$, the second at $2\omega_R$, and so on. Here, $D^*$ denotes the effective diameter, which is smaller than the actual device diameter because the EMPs propagate at different velocities in the gated and ungated regions and it depends on field and ranges from \qtyrange[]{700}{800}{\micro\meter} for the large device with geometric diameter of \qty{1225}{\micro\meter}. The derivation of $D^*$ is provided in the Supplementary Information.

It is possible to add a parasitic matrix $Y_{\rm par}$ to take into account the effects like the capacitance to ground $C_{\rm gnd}$ and the gate-gate capacitance $C_{\rm gg}$~\cite{Bosco2017,Mahoney2017}. These capacitances are expected to smaller than a $\rm pF$, having negligible effect and therefore set to zero.

Besides extracting the fit parameters the model allows us to estimate the device reflection coefficients: although it is experimentally challenging to fully calibrate the microwave lines and directly measure these quantities, the model allows us to model $\overline{S}_{11}$ and $\overline{S}_{22}$, as shown in the Supplementary Information.

We first define the impedance matching parameter
\begin{equation}
\alpha \equiv 2 Z_0 \sigma_0 ,
\label{eq:alpha_def}
\end{equation}
which relates the transmission line impedance to the internal impedance of the device~\cite{Bosco2017}.

Using this definition, the dimensionless admittance matrix $Y$ is converted into the scattering matrix $S$ via
\begin{equation}
S = \left( I/\alpha + Y \right)^{-1} \left( I/\alpha - Y \right).
\label{eq:Y_to_S}
\end{equation}

The device can show gyration for values of $\alpha<1$, and in the the experimental data in Fig.~\ref{fig:Fig5}, $\alpha\approx0.1$. Part of the parameter space is explored in the Supplementary Information.
The resulting matrix can be rescaled by the characteristic frequency \(\omega_R\) to physical units of frequency.

\subsection{Setup}
The offset estimated for the electrical delay in Fig.~\ref{fig:Fig2}a is \qty{39.2}{\nano\second}, corresponding to  $\sim$\qty{8}{\meter} cable length, consistently matching with the experimental setup.
Measurements are performed in a BLUEFORS XLD dilution refrigerator with a mixing chamber temperature of  $\sim$\qty{50}{\milli\kelvin} while measuring  with an out of plane magnetic field $B_\perp$. No  amplification is used in the measurement chain and a single RF line has an attenuation of $\sim$\qty{20}{\deci\bel} at \qty{1}{\giga\hertz}.
The measurement is performed using a Rohde-Schwarz ZNB8 vector network analyzer, with the output power set at \qty{3}{\dBm} and the measurement bandwidth set at \qty{50}{\hertz}.
\subsection{Data availability}
The data supporting the plots of this paper are available at the Zenodo
repository at xxxxx.xxxxx
\subsection{Author Contributions}
S.B. conceived the devices and adapted the theoretical framework. C.R. and W.W. designed, grew, and characterized the GaAs/AlGaAs heterostructure by transport measurements. A.T. and Y.Z. designed and fabricated the devices. A.T., Y.Z., R.S.E., and T.P. designed and executed the experiments, analyzed the data. D.M.Z. supervised the project. A.T. and D.Z. wrote the manuscript with input from all authors.

\subsection{Acknowledgments}
We thank Michael Steinacher, Sascha Linder, Sascha Martin, and Sergii Kokhas for their technical support. We are grateful to David P. DiVincenzo for insightful discussions on the theoretical aspects. We also thank Omid Sharifi Sedeh,  Miguel J. Carballido, Henok Weldeyesus, and Ilya Golokolenov for valuable discussions about the measurements and for their helpful comments on the manuscript. This research was supported by the Swiss National Science Foundation (grant no. 215757), NCCR SPIN of the SNSF (grant no. 225153), UpQuantVal InterReg. and the Swiss Nanoscience Institute.
The authors declare no conflict of interest.

\clearpage

\section*{Supplementary Information}

\setcounter{figure}{0}
\setcounter{table}{0}
\renewcommand{\thefigure}{S\arabic{figure}}
\renewcommand{\thetable}{S\arabic{table}}

\renewcommand{\thesection}{S\arabic{section}}

\section{Reverse transmission phase delay}
\begin{figure}[htpb]
\centering
\includegraphics[width=0.8\linewidth]{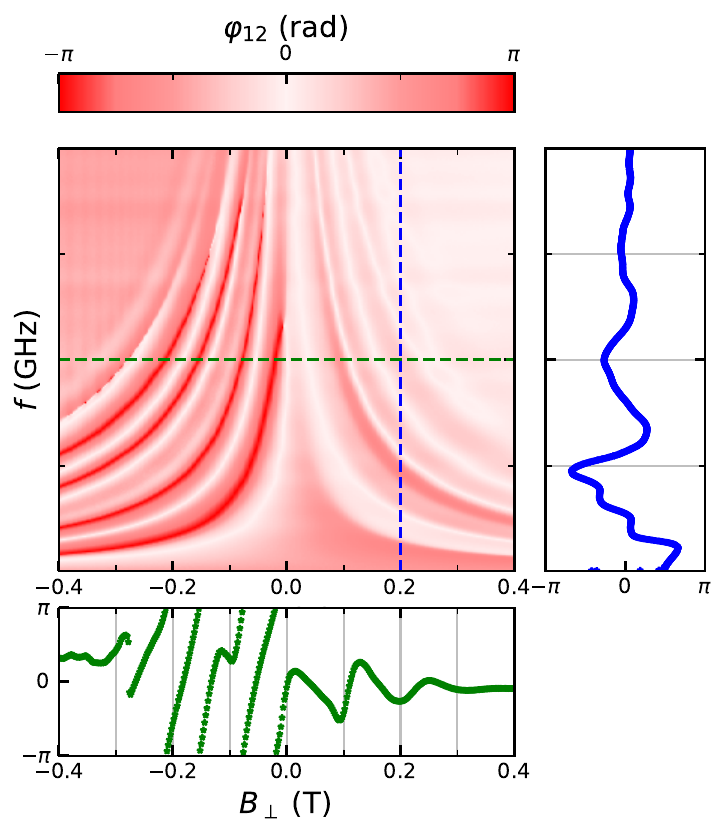}
\caption{\textbf{Reverse phase response.}
Phase of the reverse transmission parameter \( S_{12} \) of the large device, after subtraction of the electrical delay. Horizontal and vertical cuts at fixed frequency and magnetic field are shown in the bottom and side panels as indicated by the dashed lines. }
\label{fig:FigS1}
\end{figure}

In the main text of the paper we have shown the forward transmission phase by subtracting the delay of \qty{39.2}{\nano\second}, compatible with the length of the measurement setup setup of about \qty{8}{\meter}. Fig.~\ref{fig:FigS1} shows the direction not shown in Fig.~2(a) of the main text.

\section{Negative field magnitude response}
\begin{figure}[htpb]
\centering
\includegraphics[width=1.1\linewidth]{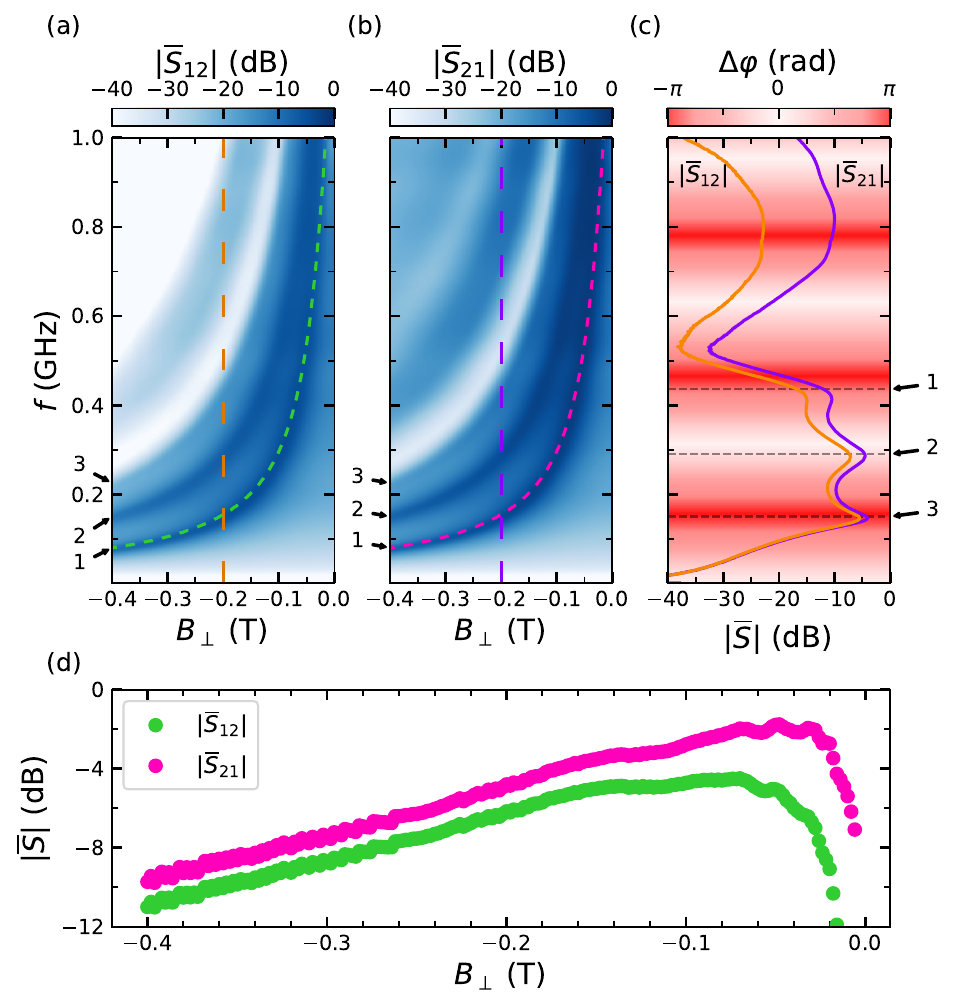}
\caption{\textbf{Magnitude response for negative fields.}
(a) and (b) Magnitude of reverse ($|\overline{S}_{12}|$) and forward ($|\overline{S}_{21}|$) transmitted signal. The green and magenta dashed curves represent the lowest-frequency gyration mode. The black arrows point at the three-peak structure. 
(c) Field cuts at \qty{-200}{\milli\tesla}, indicated by the orange and purple dashed lines in panels (a) and (b). The corresponding phase difference \( \Delta\varphi\) is shown in the background. The black arrows and black dashed lines indicate the position of the magnitude peaks.
(d) Insertion loss of the device at the lowest frequency gyration points, as indicated by the green and magenta dashed curves in panels (a) and (b). }
\label{fig:FigS2}
\end{figure}

Fig.~\ref{fig:FigS2} shows the magnitude response in negative field. Compared to Fig.~3 the intensities of $|\overline{S}_{21}|$ and  $|\overline{S}_{12}|$ are swapped, as evident from the top left areas of Fig.~\ref{fig:FigS2}(a) and Fig.~\ref{fig:FigS2}(b).

\clearpage
\section{Small device}
\begin{figure}[htpb]
\centering
\includegraphics[width=\linewidth]{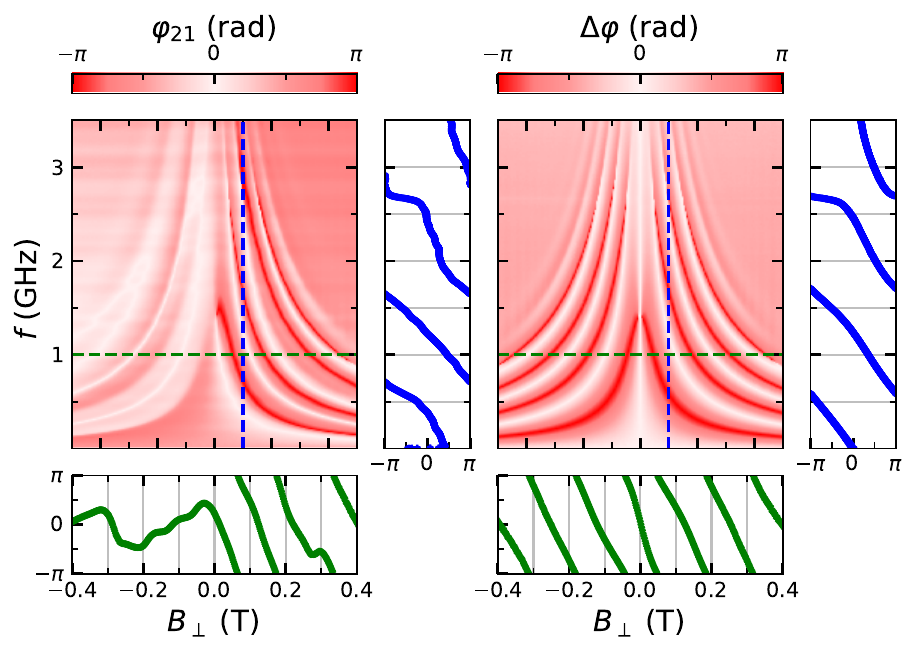}
\caption{\textbf{Phase response for the small device.}
(a) Phase of the forward transmission parameter \( S_{21} \) after subtraction of the electrical delay.
(b) Phase difference \( \Delta\varphi\) between forward and reverse transmission. Horizontal and vertical cuts at fixed frequency and magnetic field are shown in the bottom and side panels as indicated by the dashed lines.}
\label{fig:FigS3}
\end{figure}

\begin{figure}[htpb]
\centering
\includegraphics[width=\linewidth]{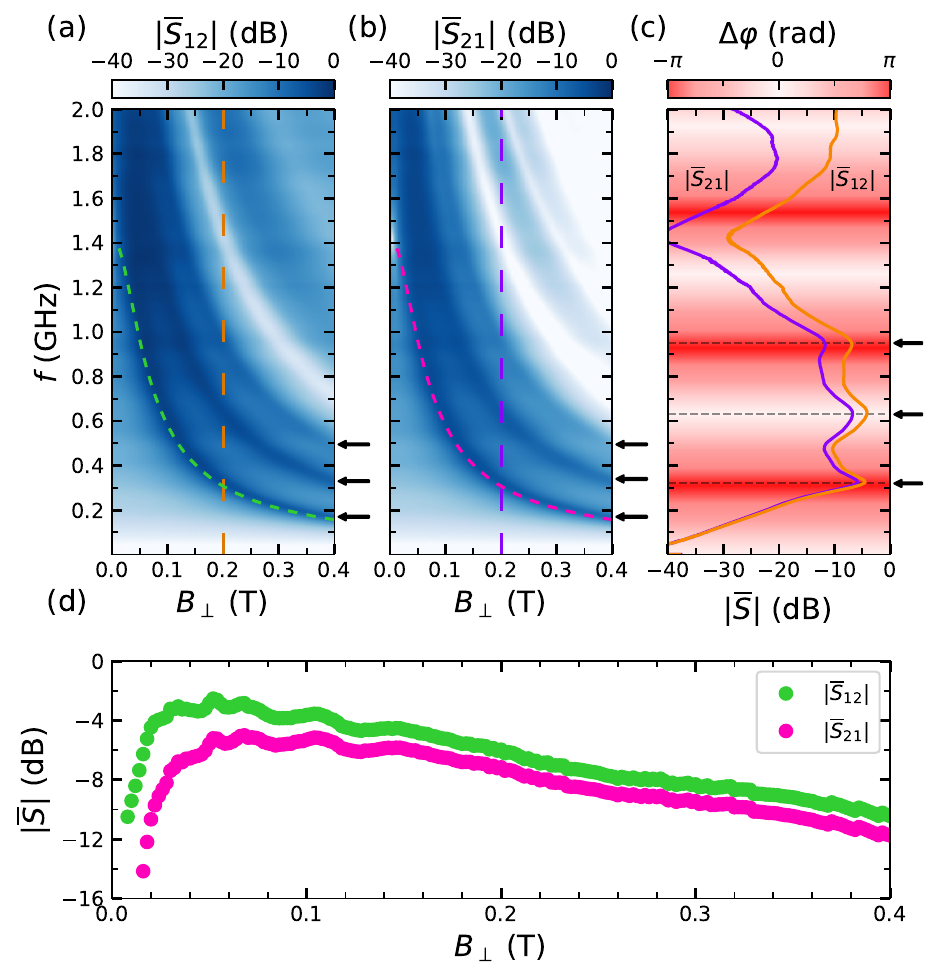}
\caption{\textbf{Magnitude response of the small device.} 
(a) and (b) Magnitude of reverse ($|\overline{S}_{12}|$) and forward ($|\overline{S}_{21}|$) transmitted signal of the small device. The green and magenta dashed curves represent lowest-frequency gyration mode. The black arrows point at the three-peak structure. 
(c) Field cuts at \qty{200}{\milli\tesla}, indicated by the orange and purple dashed lines in in panels (a) and (b). The corresponding phase difference \( \Delta\varphi\) is shown in the background. The black arrows and black dashed lines indicate the position of the magnitude peaks.
(d) Insertion loss of the device at the lowest frequency gyration points, as indicated by the green and magenta dashed curves in panels (a) and (b). }
\label{fig:FigS4}
\end{figure}

Figures Fig.~\ref{fig:FigS3} and Fig.~\ref{fig:FigS4} are analogous to Fig.~\ref{fig:Fig2} and  ~\ref{fig:Fig3} of the main paper for the smaller device. 

\clearpage 
\section{Parameters from the fit}

\begin{figure}[htpb]
\centering
\includegraphics[width=\linewidth]{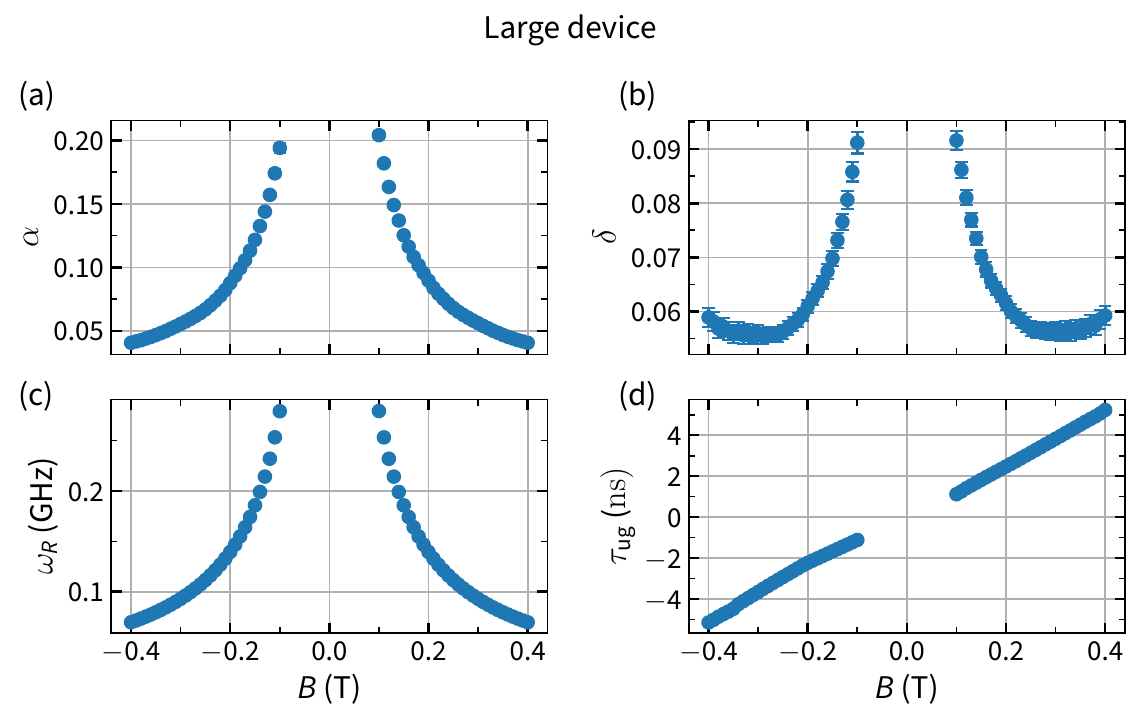}
\caption{\textbf{Fit parameters.} Extracted fit parameters for field cuts of Large device (a) Impedance matching parameter $\alpha$ (b) Dissipation $\delta$ (c) Characteristic frequency $\omega_R$ (d) Ungated time delay $\tau_{\rm ug}$.}
\label{fig:FigS5}
\end{figure}

\begin{figure}[t]
\centering
\includegraphics[width=\linewidth]{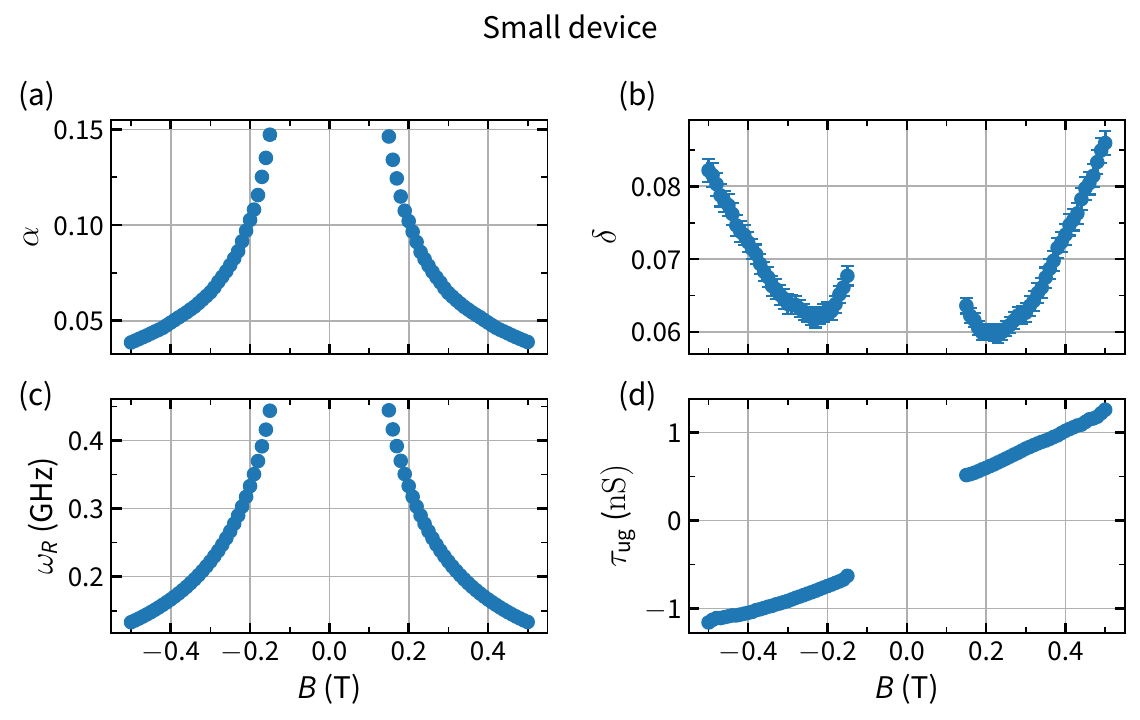}
\caption{\textbf{Fit parameters.} Extracted fit parameters for field cuts of Small device (a) Impedance matching parameter $\alpha$ (b) Dissipation $\delta$ (c) Characteristic frequency $\omega_R$ (d) Ungated time delay $\tau_{\rm ug}$.}
\label{fig:FigS6}
\end{figure}

To extract useful parameters from the data a least square curve optimization is performed. We fit linecuts at set magnetic field of the transmission magnitudes with the background removed and the phase difference. Overall the model has 4 parameters: the impedance matching $\alpha$ (defined as $2Z_0\sigma_0$) , the dissipation $\delta$, The characteristic frequency $\omega_R$, and the ungated time delay $\tau_{\rm ug}$.
We first fit the magnitude traces with $\alpha$, $\delta$ and $\omega_R$, then we estimate the ungated time delay $\tau_{\rm ug}$ by fitting the phase difference trace, since $\tau_{\rm ug}$ affects only the $\Delta\varphi$ with a linear offset. The results of these fits are reported in Fig~\ref{fig:FigS5} and Fig.~\ref{fig:FigS6} for the two device sizes.

\section{Simulation parameter space}
The three parameters that go into the model we use are $\alpha$, $\delta$ and $\omega_R$, theoretically defined as follows:
\[
\begin{aligned}
\alpha &\equiv 2 Z_0\,\sigma_0(B),\\
\omega_R &\equiv \frac{2\sigma_0(B)}{c_{\mathrm{emp}}\,D^*},\\
\delta &\equiv \frac{\pi}{2}-\arctan(\mu B).
\end{aligned}
\]
where $\mu$ is the mobility of the 2DEG.
In the actual device however we see how $\delta$ saturates and does not have a monotonic behavior in field. It is unclear why we observe this behavior and it can possibly be attributed to inefficient capacitive coupling, bulk dissipation channels and dielectric losses.

By analyzing the parameter space it will become clear that we can achieve impedance matched gyration at high fields ($\alpha\to0$) only if $\delta$ vanishes.
it is useful to turn the knobs of $\alpha$ and $\delta$ independently and leaving the scaling factor $\omega_R$ constant, for convenience set like the constant field cut of \qty{200}{\milli\tesla} in the large device that is shown in the figures of the manuscript.

\begin{figure*}[htpb]
\centering
\includegraphics[width=0.95\linewidth]{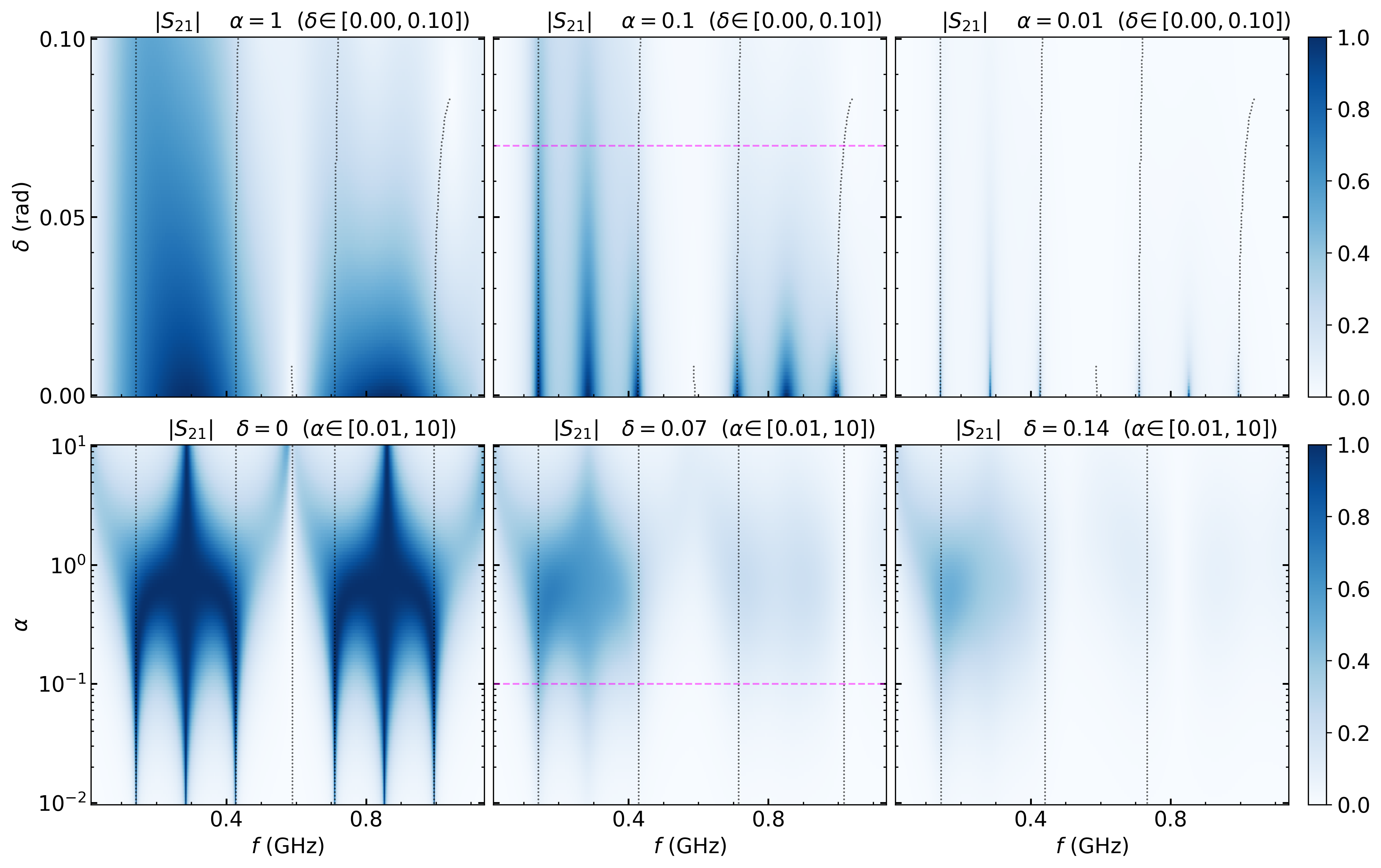}
\caption{\textbf{Exploring parameter space for transmission.}
Simulated forward transmission for different values of dissipation $\delta$ and impedance matching parameter $\alpha$. The dotted black lines represent points of gyration where ($\Delta\varphi=\pm\pi$). The dashed pink lines correspond to the extracted values from the Large device cuts shown in Fig.~\ref{fig:Fig5} of the main text. }
\label{fig:FigS7}
\end{figure*}

The bottom left panel of Fig.~\ref{fig:FigS7} shows that having $\delta=0$ does not guarantee gyration points on magnitude peaks and small values of $\alpha$ are necessary. This mismatch makes the estimation of the peak-gyration delay introduced by the ungated section $\tau_{\rm ug}$ more complex and it is sensible to first fit the magnitude data, and then adjust the phase of the model to extract $\tau_{\rm ug}$.

The most important takeaway is that smaller $\alpha$ values are not always desirable for systems with dissipation: the center top panel shows defined peaks even for larger values of $\delta$, while on the other hand in the top left panel it can be seen how the narrow peaks quickly fade as soon as dissipation is introduced.

\begin{figure*}[htpb]
\centering
\includegraphics[width=0.95\linewidth]{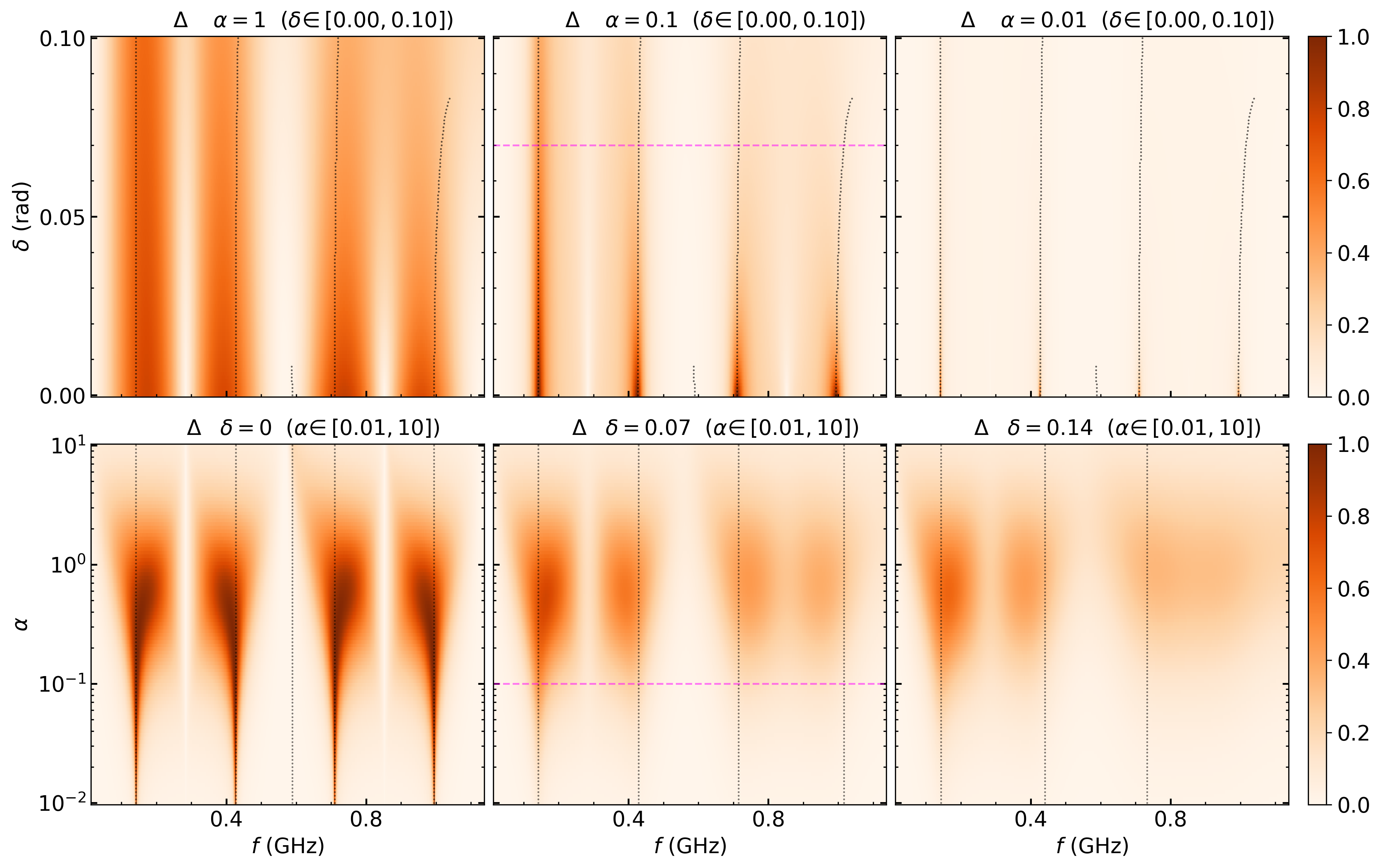}
\caption{\textbf{Exploring parameter space for performance.}
Simulated performance $\Delta$ for different values of dissipation $\delta$ and impedance matching parameter $\alpha$. The dotted black lines represent points of gyration ($\Delta\varphi=\pm\pi$). The dashed pink lines correspond to the extracted values from the Large device cuts shown in Fig.~\ref{fig:Fig5} of the main text.  }
\label{fig:FigS8}
\end{figure*}

Another way of displaying the same information is to plot the performance $\Delta$ for the same simulated dataset (Fig.~\ref{fig:FigS8}).

\clearpage
\section{Velocities, local capacitance and size}

The EMP propagation velocity in the ungated section is expected to be larger than in the gated one~\cite{Volkov1988,Kumada2011} and we can directly extract the ungated velocity by using the additional delay \(\tau_{\mathrm{ug}}\) that is accumulated on an interval of length $L$. The overall ungated path difference given the device geometry with equal spacing $L$ between the gates in fact, is $L$.
Estimating the gated speed is more complex: in the model, the frequency of the peaks is linked the angular velocity $\omega_R$ of the EMPs. We first approximate the ungated propagation time as negligible compared to the gated sections and under this assumption the effective diameter of the large device is \(D^*_0 = 4L/\pi \approx \qty{700}{\micro\meter}\). From this, we can extract the gated plasmon velocity \(v_{\mathrm{g}0}\). We can then define an initial ratio of the ungated and gated velocities \(\kappa_0\),  found to be $\sim4.5$ indicating that the three ungated sections are traversed more quickly and can therefore be treated as effectively shorter by a factor of $4.5$. This yields an updated effective diameter of  
\[
D^*_1 = (4 + 3/\kappa_0)\frac{L}{\pi}.
\]  
Iterating this procedure ten times and updating \(\kappa_i\) each step, \(\kappa\) rapidly converges to about 4, corresponding to \(D^* \approx \qty{830}{\micro\meter}\). This ratio remains basically constant for magnetic fields in the range \(\qtyrange{100}{400}{\milli\tesla}\), since at low fields both the gated and ungated EMP velocities scale as \(1/B\).

It is possible to plot the gated and ungated velocities and to extract the gate capacitance per unit length to the EMP $c_{\rm emp}$.

\begin{figure*}[t]
\centering
\includegraphics[width=\textwidth]{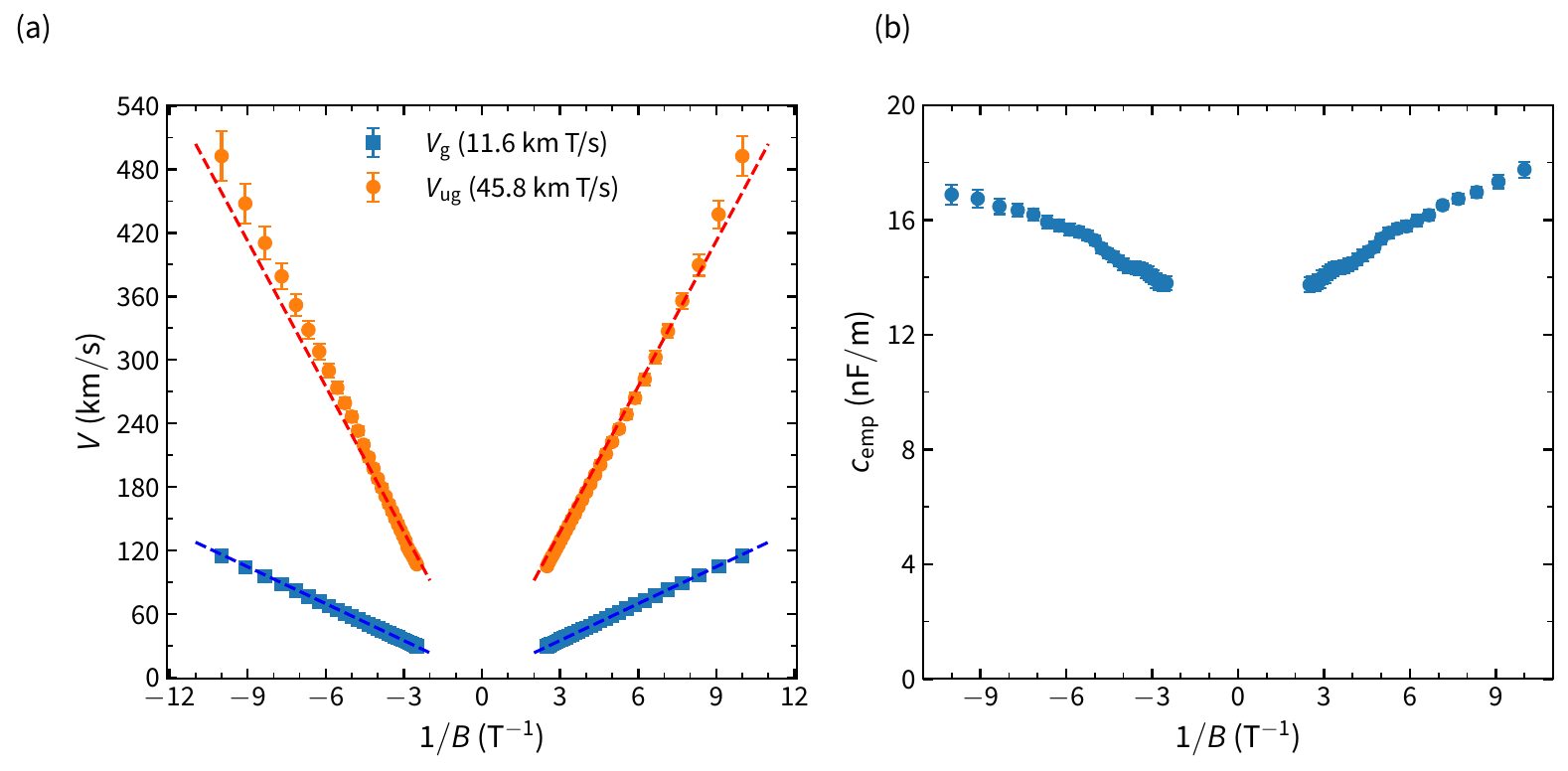}\hfill
\caption{\textbf{Large Device.} (a) Estimates of the gated and ungated EMP propagation speed for the Large device. (b) Capacitance per unit length $c_{\rm emp}$ of the EMP to the gates of width \qty{5}{\micro\meter}.}
\label{fig:FigS9}
\includegraphics[width=\textwidth]{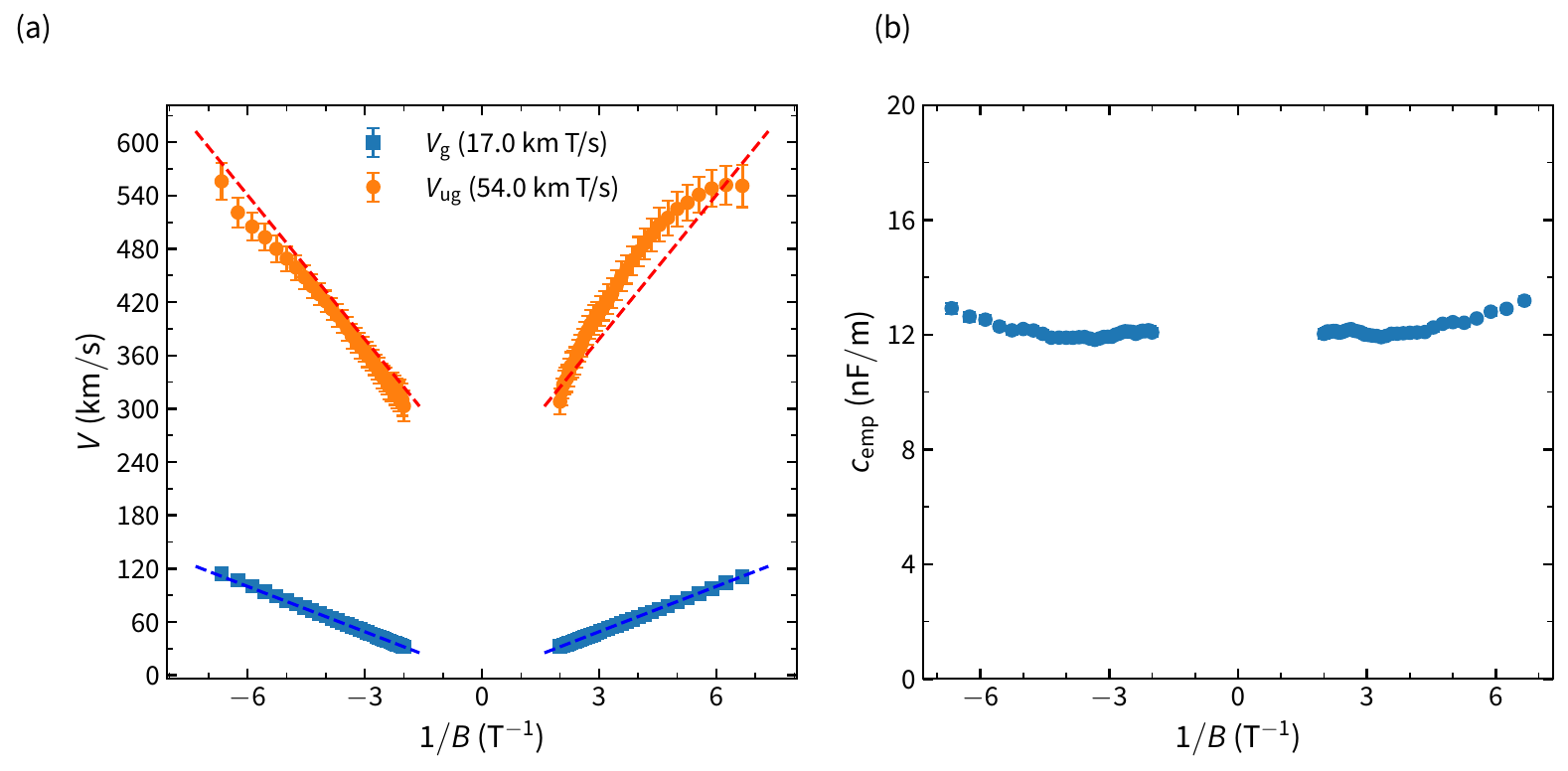}
\caption{\textbf{Small device.} (a) Estimates of the gated and ungated EMP propagation speed for the Small device. (b) Capacitance per unit length $c_{\rm emp}$ of the EMP to the gates of width \qty{3.2}{\micro\meter}.}
\label{fig:FigS10}
\end{figure*}

\section{High Field behavior}
It is natural to investigate the device at high magnetic fields. The initial expectation was that, on a quantum Hall plateau, one would observe $\delta = 0$ and $\alpha \ll 0.01$, resulting in sharp and high-amplitude transmission peaks. However, such features are not present in the measurements. This can be attributed to the finite value of $\delta$ at high fields. Features matching with integer filling factors are visible in the non-reciprocal phase difference $\Delta\varphi$, matching with transport measurements performed on a separate sample from the same wafer used for characterization. These features are reminiscent of features observed on the velocity of gated EMPs in ref.~\cite{Kumada2011}. Although some features are also visible in the transmission amplitude, the signal is comparable to the noise floor of the VNA of \qty{-110}{\deci\bel}, suggesting that the device is acting as a complete signal block. Nonetheless, improved performance may be achievable with reduced dissipation using alternative materials or interface designs, differing from capacitive coupled gates.

\begin{figure*}[htpb]
\centering
\includegraphics[width=1\linewidth]{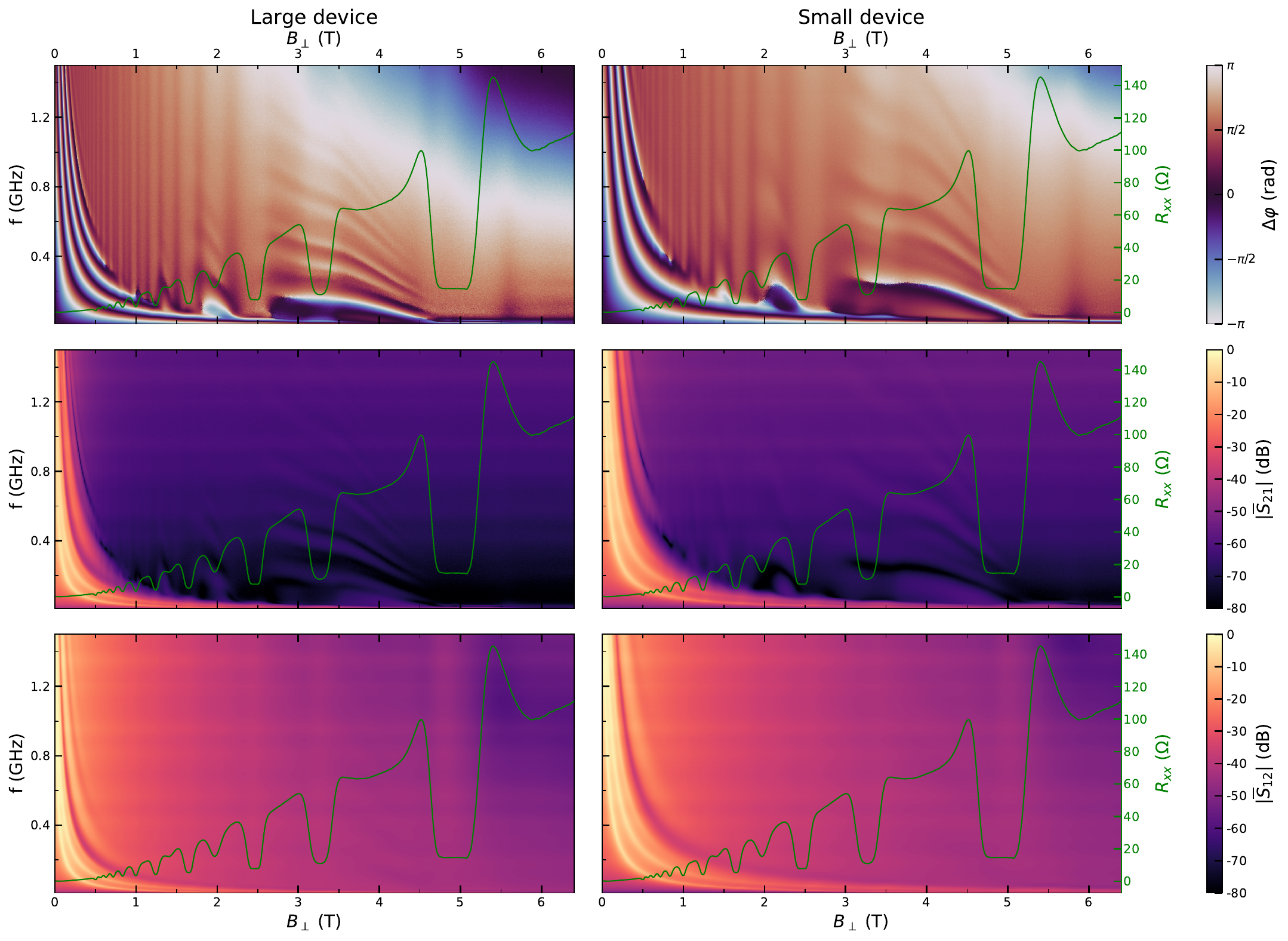}
\caption{\textbf{High field low frequency behavior.} Response of the large device (left column) and the small device (right column) at high fields. The $R_{xx}$ transport data measured in a different chip is shown in green. At $\sim\qty{5}{\tesla}$ the filling factor is $\nu=2$.}
\label{fig:FigS11}
\end{figure*}

\begin{figure*}[htpb]
\centering
\includegraphics[width=1\linewidth]{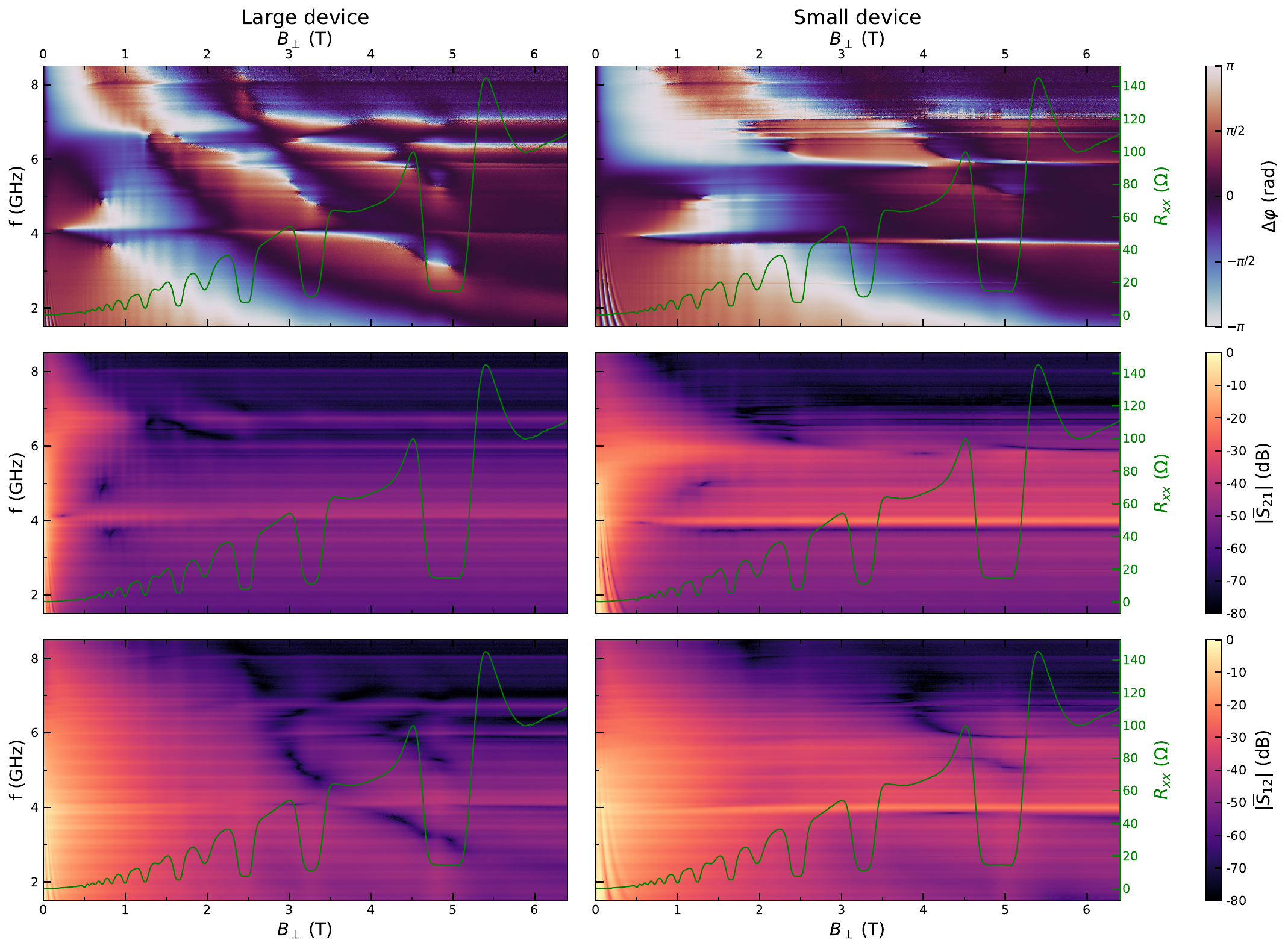}
\caption{\textbf{High field high frequency behavior.} Response of the large device (left column) and the small device (right column) at high fields and high frequencies. The $R_{xx}$ transport data measured in a different chip is shown in green. At $\sim\qty{5}{\tesla}$ the filling factor is $\nu=2$. }
\label{fig:FigS12}
\end{figure*}

\section{Reflection parameters}
\begin{figure*}[htpb]
\centering
\includegraphics[width=0.7\linewidth]{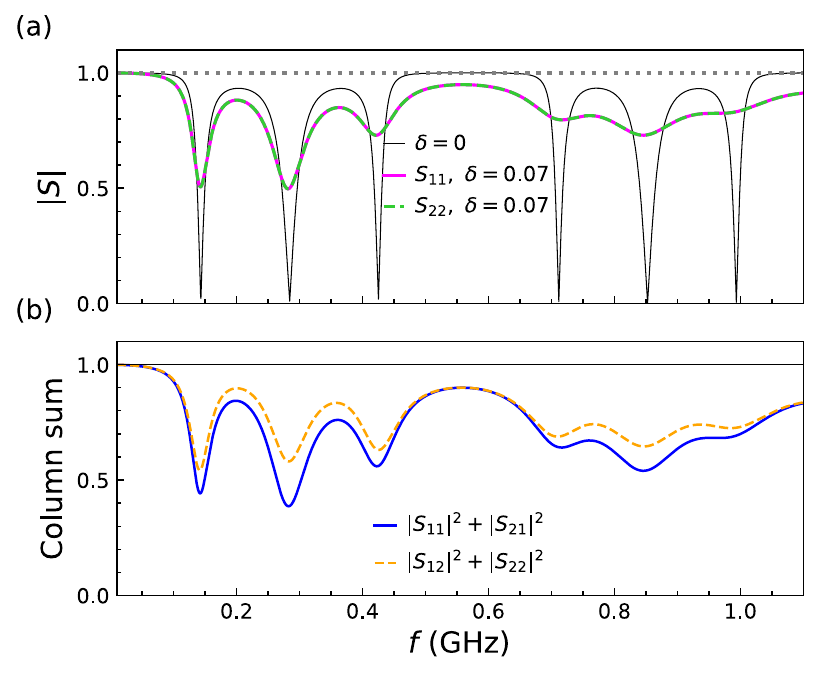}
\caption{\textbf{Reflection parameters}
(a) Reflection parameters modeled after the traces of Fig.~5(a-b) of the main text. The solid black line represents the dissipationless case $\delta=0$ while the gray dotted line represents complete reflection.
(b) Column sum of the simulated S-matrix. The solid black line is the dissipationless case $\delta=0$.
}
\label{fig:FigS13}
\end{figure*}

An Open-Short-Match (OSM) calibration is normally required to remove systematic errors from the measurement setup, such as cable losses and phase delays, and to directly access the reflection coefficients at the device plane. Since a cryogenic OSM calibration was not available in our experiment, it is not possible to directly measure $S_{11}$ and $S_{22}$. Instead, by fitting the transmission parameters with our model, we can infer the reflection coefficients, as shown in Fig.~\ref{fig:FigS13}(a). Interestingly, while the model reproduces the asymmetric transmission observed in Fig.~\ref{fig:Fig3} of the main text, it yields identical values for the two reflection coefficients. This indicates that the in the model the impedance of the stubs is the same, and that the predicted losses arise during propagation rather than at the contacts. Another useful way to see losses is to examine the unitarity of the column of the scattering matrix. In Fig.~\ref{fig:FigS5}(b), it is evident that the modeled $S$-matrix is not unitary, with larger deviations occurring near transmission magnitude peaks. This behavior is expected: when the signal couples more strongly into the device, the relative amount of dissipation increases, leading to a stronger departure from unitarity.

\clearpage

\input{ref.bbl}

\end{document}